\documentclass[12pt]{iopart}

\usepackage{graphics}
\usepackage{epsfig}
\usepackage{xspace}
\usepackage{amssymb}
\usepackage{color}

\newcommand{\PT}{{\cal PT}}

\newcommand{\rL}{{\mathrm{L}}}
\newcommand{\rR}{{\mathrm{R}}}

\newcommand{\be}{\begin{equation}}
\newcommand{\ee}{\end{equation}}

\newcommand{\sech}{\mathrm{\,sech\,}}

\begin{document}

\title[Potentials with multiple spectral singularities]{Construction of
	potentials with multiple spectral singularities}

\author{Vladimir V. Konotop$^{1,2}$ and Dmitry A. Zezyulin$^{3}$ }

\address{$^1$Centro de F\'{\i}sica Te\'orica e Computacional  
	Universidade de Lisboa, Campo Grande, Edif\'icio C8, Lisboa  1749-016, Portugal
	\\
	$^2$Departamento de F\'{\i}sica, Faculdade de Ci\^encias, Campo Grande, Edif\'icio C8, Lisboa  1749-016, Portugal\\
	$^3$ITMO University, St. Petersburg 197101, Russia 	
	 }
\ead{dzezyulin@itmo.ru}

\vspace{10pt}

\begin{abstract}
We develop an approach for designing  complex potentials with  two or  three coexisting spectral singularities in the spectra of the respective Schr\"odinger operators. The approach is illustrated with several examples. In addition, we offer a simple   recipe to create   a second-order spectral singularity by   intentionally colliding    two simple ones.
\end{abstract}

%
%
%
%
%

\section{Introduction}

Spectral singularities (SSs) in spectra of non-Hermitian operators is a long-standing subject in mathematics pioneered about  sixty years ago 
\cite{SS}. Independently, in physical literature \cite{SSphysics} of approximately the same years various effects were reported that 
are presently known   to be   originated by SSs. These are the effects of coherent perfect absorption (\emph{alias} CPA or antilasing) and lasing. About a decade ago, the interest in SSs was regained chiefly due to a series of theoretical studies~\cite{Stone,Mostafazadeh2009,Mostafazadeh2009JPA,ScarfII,Longhi}  where the relation between SSs and physical phenomena of absorption and lasing was clarified, and experiments on coherent perfect absorption of electromagnetic waves \cite{CaoScience}. For a more detailed discussion of the role that SSs play in scattering properties of physical systems see   reviews~\cite{SSreviews}.


In order to obtain a SS in a spectrum of a non-Hermitian Schr\"odinger operator, the shape of the underlying complex potential needs to  be  tuned precisely, %
i.e. by far not every   complex potential supports  a  SSs.  Therefore the presence of  SS  is a    quite demanding property that  requires a  sharp adjustment   of the physical system. At the same time, a number of complex potentials with   SSs at a given wavelength is infinite. In the one-dimensional setting, to which our discussion will be limited, such potentials 
can be constructed using different algorithms~\cite{Mostafazadeh2014,KLV}. Moreover, it has been shown in~\cite{ZezKon2020}, that all  complex potentials of the stationary one-dimensional Schr\"odinger equation with   SSs in their spectra have a universal analytical form. This last feature yields an alternative way of constructing potentials, either with  on a finite support or distributed over   the whole real axis, that feature a SS at a given wavenumber $k_0$. 

It is also known, that a spectrum of a given potential may have more than one SSs. In particular, this is the case when both $k_0$ and $-k_0$ happen to be  SSs simultaneously. This situation  is typical for   parity-time ($\PT$-)symmetric systems~\cite{Longhi}. Such SSs are often termed self-dual~\cite{rectang} and represent an especially appealing   physical situation when the same  system can operate  as a laser or a CPA for waves of the same length. Self-dual spectral singularities are also known to exist in complex potentials with other antilinear symmetries \cite{ZezKon2020,KZ2017}  and in    potentials   without $\PT$ symmetry  \cite{rectang}.  An alternative way to obtain a pair of SSs in the spectrum consists in firstly creating a single second-order SS and then splitting it into two first-order ones by a small perturbation \cite{KLV,HHK}. In this scenario, the emerging SSs will have close wavenumbers.  Some exactly solvable models with multiple spectral singularities are also known~\cite{AhmedNew,BorZez19}. At the same time,
%
%
%
there is an open general question about a \emph{systematic} approach to construction of complex potentials having two or {\em more} SSs  at \emph{arbitrary} prescribed wavenumbers. In this paper we offer a solution to   this problem. 

The rest of this paper is organized as follows.  In section~\ref{sec:gen:constr} we present a general approach to construction of potentials with several SSs. In section~\ref{sec:two} we apply the approach to construct  potentials with two coexisting SSs. In section~\ref{sec:collision} we demonstrate how our method can be used to generate a higher-order SS by colliding two first-order SSs. In section~\ref{sec:pseudo} we discuss peculiarities of the approach  
in a situation when the potential has a special antilinear symmetry, i.e. a property of pseudo-Hermiticity. In section~\ref{sec:three} we construct a potential with three spectral singularities. Section~\ref{sec:concl} concludes the paper.

\section{Construction of potentials with several spectral singularities}
\label{sec:gen:constr}

Consider a normalized one-dimensional stationary Schr\"odinger equation 
\begin{equation}
\label{SE}
-\psi''+U(x)\psi=k^2\psi
\end{equation}
on the whole axis, $x\in\mathbb{R}$, where $\psi =\psi(x)$ is a complex-valued wavefunction,  $U(x)$ is   a complex-valued potential  vanishing at the infinity: $\lim_{x\to\pm\infty}U(x)=0$,
$k$ is a spectral parameter (dimensionless wavenumber), and a prime  stands  for the derivative with respect to $x$. The starting point of our approach  is the  sufficient condition for the potential $U(x)$ to have a spectral singularity 
\cite[section~3]{ZezKon2020}.  Assume that the potential $U(x)$ admits the representation 
\begin{equation}
\label{Wadati}
U(x)=-w_0^2(x)-iw_0'(x)+k_0^2,
\end{equation}
where $k_0$ is a nonzero real, and the base function $w_0(x)$ has an  asymptotic behavior
\begin{eqnarray}
w_0(x) = k_0 + w_{-\infty}(x), \quad w_{-\infty}=O(|x|^{-2}) \quad \mbox{at\ } x\to-\infty,\\
w_0(x) = -k_0 + w_{\infty}(x), \quad w_{\infty}=O(|x|^{-2}) \quad \mbox{at\ } x\to\infty.
\end{eqnarray}
Then the potential $U(x)$ has a SS at  $k=k_0$, and the solution of (\ref{SE}) associated with this SS  (i.e., SS-solution) reads \cite{ZezKon2020}
\begin{equation}
\label{SSsolut}
\psi_0(x)=\rho\exp\left[-i\int_{x_0}^xw_0(\xi)d\xi\right], 
\end{equation}
where $x_0$ is any point of  the real axis, and  $\rho$ is a nonzero constant.  Indeed, it is easy to check that there exist nonzero constants $\rho_\pm$ such that
$\lim_{x\to\pm\infty}[\psi_0(x)-\rho_\pm e^{\pm ik_0 x}]=0$,
which means that  solution $\psi_0$ satisfies    purely outgoing (for $k_0>0$) or purely incoming  (for $k_0<0$)  wave  boundary conditions, i.e. corresponds to the lasing mode or to the CPA-operation (antilaser), respectively. 

Equations (\ref{Wadati}) and  (\ref{SSsolut}) presuppose that the function $w_0(x)$ is sufficiently well-behaved. However, the discussion can be generalized easily on the situation when $w_0(x)$ has a  finite number of discontinuities at points $x_1<x_2<\ldots <x_N$, provided that  $w_0(x)$ behaves as $w_0(x)=i (x-x_j)^{-1} + O(1)_{x\to x_j}$  around each discontinuity  $x_j$. In this case, instead of (\ref{SSsolut}), the SS-solution $\psi_0(x)$ should be defined piecewise on each continuity interval of $w_0(x)$, and  it is  by definition  equal to zero  at the discontinuities: $\psi_0(x_j)=0$ (see \cite{ZezKon2020} for the details). 

Suppose now that  there exist $n$ different base functions $w_1(x),\ldots , w_n(x)$, such that 
\begin{eqnarray}
\label{eq:wjlim1}
w_j(x) = k_j + w_{-\infty, j}(x), \quad w_{-\infty,j}=O(|x|^{-2}) \quad \mbox{at\ } x\to-\infty,\\
\label{eq:wjlim2}
w_j(x) = -k_j + w_{\infty, j}(x), \quad w_{\infty,j}=O(|x|^{-2}) \quad \mbox{at\ } x\to\infty,
\end{eqnarray}
where all reals  $\{k_1, k_2,\ldots, k_n\}$ are different,  and using these functions the potential  $U(x)$  can be represented in $n$ different ways:
\begin{eqnarray}
\label{Uw}
U(x)=-w_j^2(x)-iw_j^\prime(x)+k_j^2 \quad \mbox{for $j=1,2,\ldots, n$}.
\end{eqnarray}
In this situation, the potential $U(x)$ obviously features $n$ \emph{coexisting spectral singularities}  at wavenumbers $k_j$, $j=1,2,\ldots, n$. Hence, in order to find a potential with multiple spectral singularities, it is sufficient to find 
any function $w_j(x)$ in equation~(\ref{Uw}).
 However, given one of the functions $w_j(x)$, one has only the SS $k_j$ given, while determination the others requires finding {\em all} base functions $w_j(x)$ yielding (\ref{Uw}). 

Naturally,   by far not  every potential $U(x)$  can be  represented in the form (\ref{Uw}), and thus the functions $w_j(x)$, $j=1,\ldots, n$ must be interrelated. The requirement for two different base functions, say  $w_j$ and $w_{j+1}$, to generate the same potential $U(x)$ in the explicit form reads
\begin{equation}
\label{rel1}
w_{j+1}^2(x)+iw_{j+1}^\prime(x)-k_{j+1}^2=w_j^2(x)+iw_j^\prime(x)-k_j^2.
\end{equation}
For existence of $n$ SSs, this relation  must be satisfied for all $j=1,...,n-1$.

Let us look for a sequence of functions $w_j(x)$, which is generated by the recurrence relation
\begin{eqnarray}
\label{jj+1}
w_{j+1}(x)=w_j(x)+\chi_j(x), \quad j=1,\ldots, n-1,
\end{eqnarray}
where the newly introduced functions $\chi_j(x)$ satisfy the asymptotic behavior
\begin{equation}
\label{assimpt_chi}
\lim_{x\to\pm\infty}\chi_j(x)= \mp (k_{j+1}-k_{j}),\quad \lim_{x\to\pm\infty}\chi_j^\prime(x)=0,   
\end{equation}
which follows from (\ref{eq:wjlim1})-(\ref{eq:wjlim2}). For the sake of simplicity, in what follows we assume that all asymptotic values at infinities are approached fast enough, i.e. faster than $O(|x|^{-2})$. 
Using (\ref{jj+1}) and (\ref{rel1}) one can obtain two relations 
\begin{eqnarray}
i\chi_j^\prime+\chi_j^2+2w_j\chi_j+k_j^2-k_{j+1}^2=0,
\\
i\chi_j^\prime-\chi_j^2+2w_{j+1}\chi_j+k_j^2-k_{j+1}^2=0
\end{eqnarray}
from which $w_j(x)$ and $w_{j+1}(x)$ can be expressed through $\chi_j(x)$:
\begin{eqnarray}
\label{eq:ww1}
w_j=-\frac{\chi_j}{2}-\frac{i\chi_j^\prime+k_j^2-k_{j+1}^2}{2\chi_j},
\\
\label{eq:ww2}
w_{j+1}=\frac{\chi_j}{2}-\frac{i\chi_j^\prime+k_j^2-k_{j+1}^2}{2\chi_j}.
\end{eqnarray}

The obtained relations (\ref{eq:ww1}) and (\ref{eq:ww2}) already solve the problem of finding a   potential with {\em two spectral singularities}, i.e. for the case $n=2$ and $j=1$. Indeed, by choosing any   function $\chi_1(x)$ which has no real zeros,  is two times 
differentiable,  and satisfies the asymptotic requirements  (\ref{assimpt_chi}) with the prescribed SSs $k_1$ and $k_2$, one readily  obtains functions $w_{1}(x)$ and $w_{2}(x)$, and hence the potential $U(x)$. In fact, the formulated requirements on $\chi_1(x)$ can be weakened, and $\chi_1$ may be allowed to have a real zero $x_0$, provided that either $\chi_1'(x_0)=i(k_1^2-k_2^2)$ or  $\chi_1'(x_0)=i(k_1^2-k_2^2)/3$. In the former case, the singularity of functions $w_{1,2}$ at $x_0$ is removable, since using the  l'H\^opital's  rule one computes   $\lim_{x\to x_0} w_{1,2} = \chi_1^{''}(x_0)/(2(k_2^2-k_1^2))$.  In the latter case,  both functions $w_{1,2}$ have singularities of the form $i/(x-x_0)$. Nevertheless, the potential $U(x)$ remains well-behaved around $x_0$, because singular terms in  $w_{j}^2$ and $w_{j}'$ cancel each other (here $j=1,2$).  As shown in \cite{ZezKon2020} and mentioned above in this section, a singularity of function $w_j$  at $x_0$   corresponds to a  zero of the corresponding SS solution: $\psi_j(x_0)=0$.  

Additionally, the potential $U(x)$ remains well-behaved if the  function $\chi_1(x)$  has a singularity of the form $\pm i/(x-x_0)$, where $x_0$ is any point of the real axis. More precisely, if $\chi_1$ has a singularity $i/(x-x_0)$, then function $w_1(x)$ has no singularity at  $x_0$, while $w_2$ diverges as $i/(x-x_0)$. Therefore, the  SS-solution corresponding to the  wavevenumber $k_2$ has a zero at $x=x_0$. \emph{Vice versa}, if    $\chi_1$ has a singularity $-i/(x-x_0)$, then the function $w_1(x)$ diverges as $i/(x-x_0)$, and the SS-solution with the wavevenumber $k_1$ has a zero at $x=x_0$, while the function $w_2(x)$ is well-behaved at $x_0$, and the SS solution corresponding to the wavevenumber $k_2$ is zero-free (provided that $x_0$ is the only singularity of $\chi_1$).  The conditions on $\chi_j$ can be weakened further, by allowing multiple discontinuities of the first derivative on a finite set of isolated points of the real axis,  which 
ensures weaker constraints imposed on the base functions \cite{ZezKon2020}. 

If the number of SSs is {\em more  than two}, the functions $\chi_j$ are not arbitrary, but must satisfy the following relations
\begin{equation}
\label{eq:chi}
\chi_j-\frac{i\chi_j^\prime+k_j^2-k_{j+1}^2}{\chi_j}=-\chi_{j+1}- \frac{i\chi_{j+1}^\prime+k_{j+1}^2-k_{j+2}^2}{\chi_{j+1}}.
\end{equation}

If two such functions $\chi_{j}$ and $\chi_{j+1}$ are found, then one can construct a potential $U(x)$ featuring three SSs, since now $w_j(x)$,$w_{j+1}(x)$ and $w_{j+2}(x)$ and obtained self-consistently. Thus our next step is finding the functions $\chi_{j}$ and $\chi_{j+1}$. To this end we introduce  new functions $\nu_j(x)$ for $j=1,..., n-2$ by the relations 
\begin{eqnarray}
\label{nu}
\chi_{j+1}(x)=\chi_j(x)/\nu_j(x), 
\end{eqnarray}
and by the asymptotic behavior
\begin{equation}
\label{asympt_nu}
\lim_{x\to\pm\infty}\nu_j(x) =
\frac{k_{j}-k_{j+1}}{k_{j+1}-k_{j+2}}.
\end{equation}
From (\ref{eq:chi}) we obtain
\begin{equation}
\label{eq:nu_chi}
i\chi_{j}\nu_j^\prime+(k_{j+1}^2-k_{j+2}^2)\nu_j^2+(k_{j+1}^2-k_{j}^2+\chi_j^2)\nu_j
+\chi_j^2=0.
\end{equation}
Thus by choosing  some  $\nu_j(x)$ 
satisfying the asymptotic behavior (\ref{asympt_nu}), one can, in principle, determine $\chi_{j}(x)$ by solving the quadratic  equation (\ref{eq:nu_chi}).
Due to the definition (\ref{nu}), this readily gives $\chi_{j+1}$, and subsequently the base functions $w_j$, $w_{j+1}$, and $w_{j+2}$. Any of these functions can be used to obtain the complex potential $U(x)$. This resolves the problem of constructing complex potentials having {\em three spectral singularities}.

For constructing the potentials with {\em four  or  more spectral singularities}, one has to continue the self-consistent procedure described above, and ensure that $\nu_{j+1}$ and $\nu_j$ are chosen such that the function $\chi_{j+1}$ obtained from (\ref{eq:nu_chi}) with $j\to j+1$, i.e. from the equation
\begin{equation}
\label{eq:nu_chi_1}
i\chi_{j+1}\nu_{j+1}^\prime+(k_{j+2}^2-k_{j+3}^2)\nu_{j+1}^2 
+(k_{j+2}^2-k_{j+1}^2+\chi_{j+1}^2)\nu_{j+1} +\chi_{j+1}^2=0
\end{equation}
coincides with that   given by (\ref{nu}). To this end one has to find $\nu_j$ and $\nu_{j+1}$ solving the system (\ref{eq:nu_chi}), (\ref{eq:nu_chi_1}), and (\ref{nu}). So far we have not been able  to solve this problem, i.e.  to find   an algorithm of constructing Schr\"odinger operators with more than three SSs (which of course does not mean nonexistence of the respective complex potentials). Therefore now we turn to illustration of our approach   considering potentials with    two and three coexisting SSs. 

\section{Two spectral singularities}
\label{sec:two}

\subsection{Self-dual spectral singularity}
\label{sec:sdss}

Starting with self-dual SSs,   from (\ref{eq:ww1}) with $j=1$  and $k_1=-k_2$ we obtain 
\begin{equation}
w_1= -\frac{\chi_1^2+i\chi_1'}{2\chi_1},
\end{equation}
and in terms of the function $\chi_1$ the searched   potential reads
\begin{equation}
\label{eq:Uchi}
U(x) = \frac{3(\chi_1')^2-2\chi_1\chi_1''-\chi_1^4}{4\chi_1^2} + k_1^2.
\end{equation}
Simple symmetry considerations show that if $\chi_1(x)$  is an  anti-parity-time (anti-$\PT$)-symmetric function, $\chi_1(x) = -\chi_1^*(-x)$, then the potential $U(x)$ is $\PT$ symmetric, i.e.  $U(x)=U^*(-x)$~\cite{BenderBoet}. This recovers a known property of  $\PT$-symmetric potentials to feature self-dual SSs~\cite{Longhi}.  In a more general case of the function $\chi_{1}(x)$ not being anti-$\PT$-symmetric, using (\ref{eq:Uchi}) one can systematically obtain  asymmetric potentials $U(x)$ with  self-dual SSs  for any $k_1$ given beforehand (particular examples of 
asymmetric potentials with self-dual SSs have been previously 
discussed in~\cite{ZezKon2020,rectang,KZ2017}).   

Using this approach, one can  easily construct complex potentials featuring two SSs, $n=2$, having several free parameters, i.e.  being deformable.  We illustrate this showing the potentials depending on two complex parameters $a_{0,1}$. As the generating function $\chi_1(x)$ we use
\begin{equation}
\label{chi1}
\chi_1(x) = (k_1-k_2)\tanh x  + ia_0\sech(x-a_1), 
\end{equation}
where $a_{0} \ne 0$ and $a_1$ are  free parameters. The resulting potential $U(x)$ is $\PT$ symmetric if $a_0$ is real and $a_1=0$ and is asymmetric otherwise. Using computer algebra, for $k_2=-k_1$ we obtain the resulting potential in the form
\begin{eqnarray}
\label{example1}
\nonumber
U(x)=k_1^2+\{ 
k_1^2\sech^4x(3 + 4\sinh^2x - 4k_1^2\sinh^4x)
\\
\nonumber
+ia_0k_1\sech y[2\tanh x \sech^2 y - 3\tanh y \sech^2x
\\ \nonumber
-\tanh x(8k_1^2\tanh^2x+1-2\sech^2x)]\\
\nonumber
-(a_0^2/4)\sech^2y(1+\sech^2y - 24k_1^2\tanh^2x)\\
+ 2i a_0^3  k_1\tanh x\sech^3 y - (a_0^4/4)\sech^4 y
\} /\chi_1^2(x),
\end{eqnarray}
where  for the sake of brevity we have introduced  $y=x-a_1$. 

In order to {illustrate the above results using examples of the  scattering data of the constructed potentials, we employ} the   transfer matrix formalism, introducing a pair of left (superscript ``$\rL$'') and right (superscript ``$\rR$'') Jost solutions  which for real $k$  are defined uniquely by their asymptotic behaviors   
\begin{eqnarray}
\label{Jost}
\begin{array}{ccc}
\phi_1^{\rL}(x;k)\to e^{ikx},\qquad & \phi_2^{\rL}(x;k)\to e^{-ikx} & \mbox{\quad at $x\to-\infty$,} 
\\[2mm]
\phi_1^{\rR}(x;k)\to e^{ikx}, \qquad & \phi_2^{\rR}(x;k)\to e^{-ikx} & \mbox{\quad at $x\to+\infty$.} 
\end{array}
\end{eqnarray}
Since a scattering solution $\psi(x)$ of the Schr\"odinger equation (\ref{SE}) is 
 a linear combination of the Jost solutions:
$\psi(x) = a^\rL \phi_1^\rL + b^\rL \phi_2^\rL = a^\rR \phi_1^\rR + b^\rR \phi_2^\rR$,
one obtains the relation between the coefficients $a^{\rL,\rR}$, $b^{\rL,\rR}$ defining the  
 transfer matrix $M(k)$:
\begin{equation}
\left(\begin{array}{c}
a^\rR\\b_\rR
\end{array}\right) = M(k) \left(\begin{array}{c}
a^\rL\\b_\rL
\end{array}\right), \qquad M(k)= \left(\begin{array}{cc}
M_{11}(k)&M_{12}(k)\\
M_{21}(k)&M_{22}(k)
\end{array}\right).
\end{equation}
%
Real zeros of the  matrix element $M_{22}(k)$ are SSs. Left and right transmission and reflection coefficients are computed as
\begin{equation}
 T^\rL = T^\rR=T = \frac{1}{M_{22}}, \qquad R^\rL = -\frac{M_{21}}{M_{22}}, \qquad R^\rR = \frac{M_{12}}{M_{22}}.
\end{equation}
Below we consider the scattering data $T(k)$, $R^\rL(k)$ and $ R^\rR(k)$ formally for both, positive and negative wavenumbers $k$, which naturally affects their physical interpretation (which is more conventionally is applied to positive $k$). A positive SS  $k>0$  is associated with 
 a  lasing solution, while  a  negative SS    $k<0$   describes  a CPA  solutions.

\begin{figure}
	\centering
	\includegraphics[width=0.99\columnwidth]{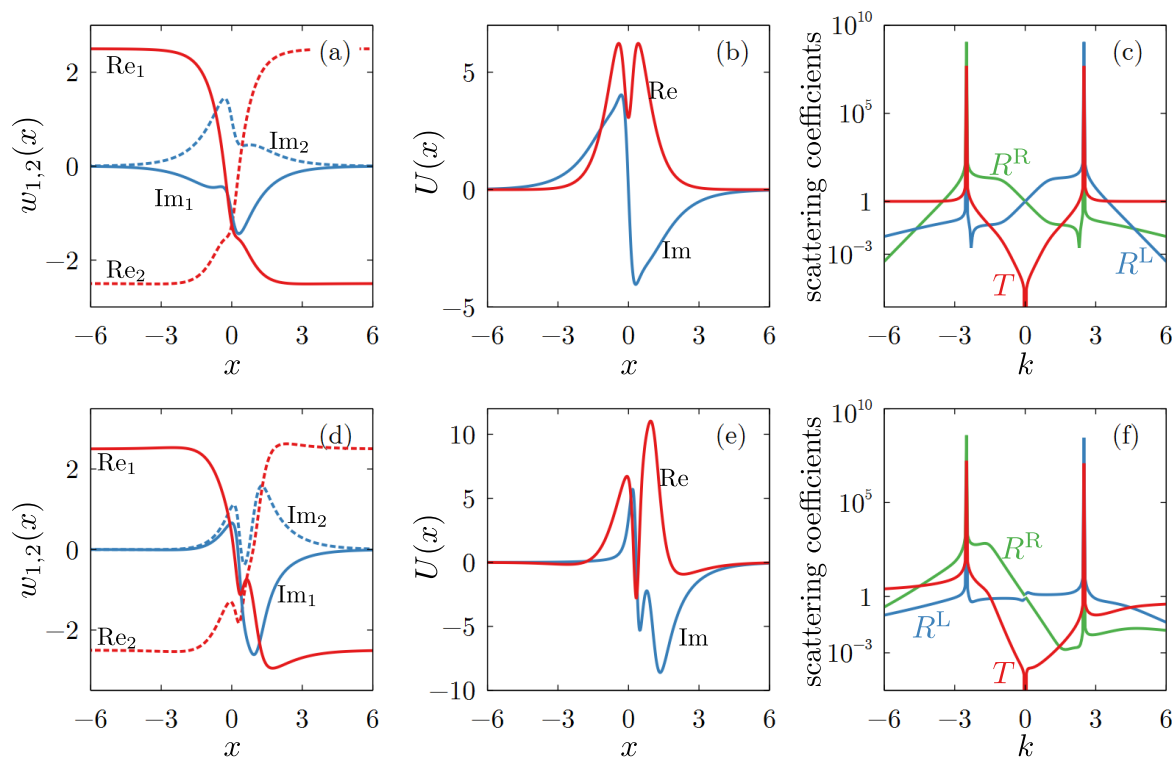}
	\caption{Examples of   { $\PT$-symmetric at $a_0 = 2$, $a_1 = 0$ (a--c) and asymmetric at $a_0 = 2+i$, $a_1 = 1-i$ (d--f)  potentials  (\ref{example1})} with a self-dual SS at $k_1 = -k_2 =  2.5$. 
		Panels (a,d) show real and imaginary parts of the functions $w_1(x)$ (solid lined) and $w_2(x)$ (dashed lines). Panels (b,e) show real and imaginary parts of potentials $U(x)$.  {In (a), (b), (e) and (d) the red and blue curves are the real (Re) and imaginary (Im) parts, respectively. } Panels (c,f) show  moduli of the transmission coefficient $T$ (red curves) and left and right reflection coefficients $R^{\mathrm{L},\mathrm{R}}$  (blue and green curves, respectively).}
	\label{fig:sdss} 
\end{figure}

Once the potential $U(x)$ is given in an analytical form, the transfer matrix and the scattering characteristics can be computed numerically. In figure~\ref{fig:sdss} we  illustrate the shape of the potential by (\ref{example1}) for two combinations of free parameters $a_{0,1}$ which correspond to the $\PT$-symmetric case [figure~\ref{fig:sdss}(a,b)] and to the asymmetric case [figure~\ref{fig:sdss}(d,e)] and for $k_1=-k_2=2.5$ in either case. In the right column we show  the numerically computed scattering characteristics for the respective potentials [figure~\ref{fig:sdss} (c) and (f) for symmetric and asymmetric cases]. In accordance with the analytical results, the moduli of  scattering coefficients for both potentials feature two extremely sharp peaks at $k_1$ and $k_2$, illustrating self-duality of the SS.

\subsection{Two independent spectral singularities}

\begin{figure}
	\centering
	\includegraphics[width=0.99\columnwidth]{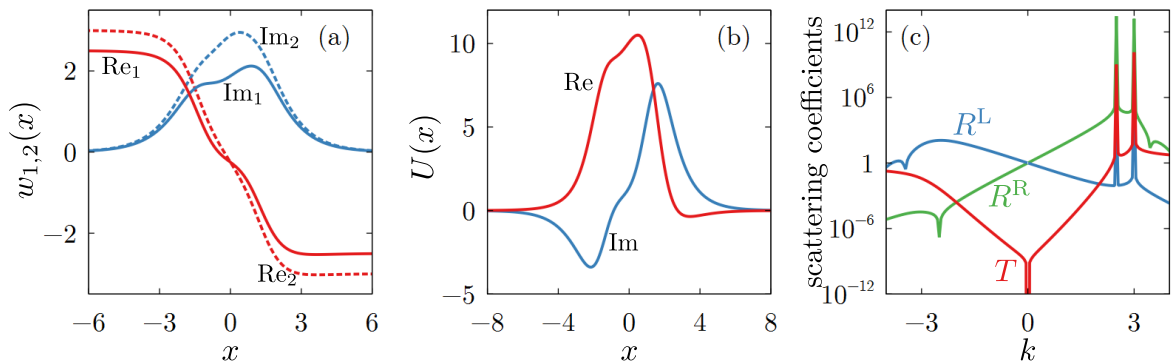}
	\caption{Example of a potential with two independent SSs at $k_1=2.5$ and $k_2=3$. The functions $w_{1,2}$ and the potential $U(x)$ are obtained using (\ref{chi1}) 
		with $a_0=1$ and $a_1=0$. The panels are organized in the same way as those in  figure~\ref{fig:sdss}(a,b,c). }
	\label{fig:2ss} 
\end{figure}

Now we dismiss the requirement of self-duality and present an example of a  complex potential with  spectral singularities at \emph{any} two  wavevectors $k_{1,2}$ chosen beforehand.  We use   the two-parametric family given by equation~(\ref{chi1}) with $k_1\ne -k_2$.  Substituting (\ref{chi1})   into      (\ref{eq:ww1}) and (\ref{eq:ww2}) with $j=1$, one can find explicitly the base functions $w_{1,2}(x)$ and recover the potential $U(x)$ in the analytical form. Since the resulting analytical    expressions are rather bulky, we do not present them in the text and limit the presentation by plotting functions $w_{1,2}$ and potential $U(x)$ in figure~\ref{fig:2ss}. This figure illustrates the situation when the potential lases at two different wavevenumbers: $k_1=2.5$ and $k_2=3$. Notice that the constructed SSs are not self-dual, i.e. the corresponding scattering coefficients peaks do not have counterparts in the negative $k$ half-axis.   Potentials with spectral singularities at any other combinations of wavevenumbers can be also constructed easily.

\begin{figure}
	\centering
	\includegraphics[width=0.99\columnwidth]{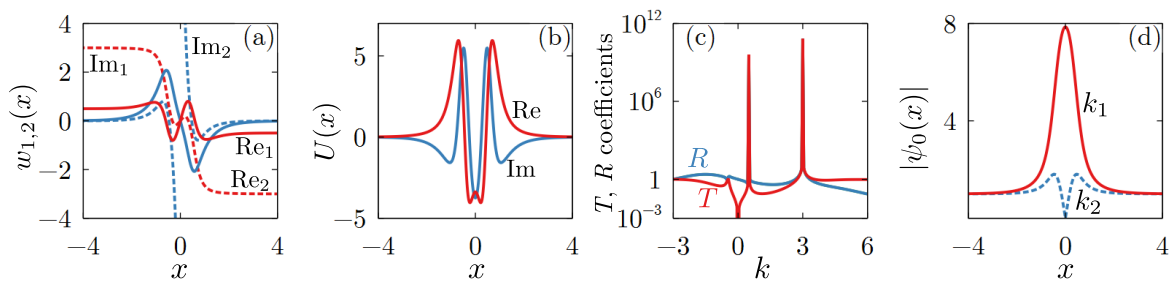}
	\caption{Example of a potential with two independent SSs  at $k_1=0.5$ and $k_2=3$ one of which corresponds to a SS-solution with a zero. Functions $w_{1,2}$ and the potential $U(x)$ are obtained using (\ref{chi-sing}). Panels (a-c) are organized in the same way as those in figure~\ref{fig:2ss} (the potential is even function, hence moduli of left and right reflection coefficients are equal), and (d) shows amplitudes of SS-solutions associated with $k_1$ and $k_2$  (solid red line  and  dotted blue line, respectively).}
	\label{fig:node} 
\end{figure}

In the above examples the SS-solutions do not vanish on the real axis. In order to construct a potential enabling a SS solution with a real zero, we explore the function $\chi_1(x)$ having a singularity at $x=0$: 
\begin{equation}
 \label{chi-sing}
 \chi_1(x)=(k_1-k_2)\tanh x + i\frac{\sech x}{x}.
\end{equation}
The respective complex potential $U(x)$ is a symmetric (even) function. A representative example of such a potential is shown in figure~\ref{fig:node}. In agreement with the discussion in section~\ref{sec:gen:constr}, the SS solution associated with the wavenumber $k_1$ is zero-free, whereas that with $k_2$ has a zero at $x=0$. Without loss of generality, both shown SS solutions plotted in Fig.~\ref{fig:node}(d,e) are normalized such that $\lim_{x\to \pm\infty}|\psi|=1$.


\section{A  {second}-order spectral singularity from the collision of two first-order ones}
\label{sec:collision}

Our approach to construction of multiple coexisting SSs  proves to be   useful also for designing potentials with a higher-order SS.  In this section we solve the problem  of creating a second-order SS by intentionally colliding two simple SSs $k_1$ and $k_2$ (i.e. implementing the mechanism inverse to the splitting of a second-order SS considered previously~\cite{KLV,HHK}).  To this end, we rewrite relations (\ref{eq:ww1})-(\ref{eq:ww2}) with $j=1$, but make the dependence on $k_{1,2}$ explicit:
\begin{eqnarray}
\label{eq:ww1k}
w_1(x;k_1,k_2)=-\frac{\chi_1(x;k_1,k_2)}{2}-\frac{\displaystyle i\frac{\partial\chi_1(x; k_1,k_2)}{\partial x}+k_1^2-k_{2}^2}{2\chi_1(x;k_1,k_2)},
\\[3mm]
\label{eq:ww2k}
w_{2}(x;k_1,k_2)=\frac{\chi_1(x;k_1,k_2)}{2}-\frac{\displaystyle i\frac{\partial\chi_1(x; k_1,k_2)}{\partial x}+k_1^2-k_{2}^2}{2\chi_1(x;k_1,k_2)}.
\end{eqnarray}
Let $k_1$ be fixed  and $k_2$ approach  $k_1$.  We look for the situation when two SSs collide and result in the identical SS solution, i.e.,  $\lim_{k_2\to k_1}w_1 = \lim_{k_2\to k_1}w_2$. This is possible   if $\lim_{k_2\to k_1}\chi_1=\lim_{k_2\to k_1}\frac{\partial \chi_1}{\partial x} = 0$.  Then we can use the l'H\^opital's rule to compute
\begin{equation}
\label{eq:tw}
\tilde{w}(x):=\lim_{k_2\to k_1}w_1 = \lim_{k_2\to k_1}w_2 = -\lim_{k_2\to k_1}\frac{\displaystyle i\frac{\partial^2\chi_1(x; k_1,k_2)}{\partial k_2\partial x} - 2k_2}{2\displaystyle \frac{\partial\chi_1(x; k_1,k_2)}{\partial k_2}}.
\end{equation}
If the latter limit exists, then the potential $\tilde{U}(x) = -\tilde{w}^2-i\tilde{w}' + k_1^2$ is expected to feature a  second-order SS.

\begin{figure*}
	\centering
	\includegraphics[width=0.99\textwidth]{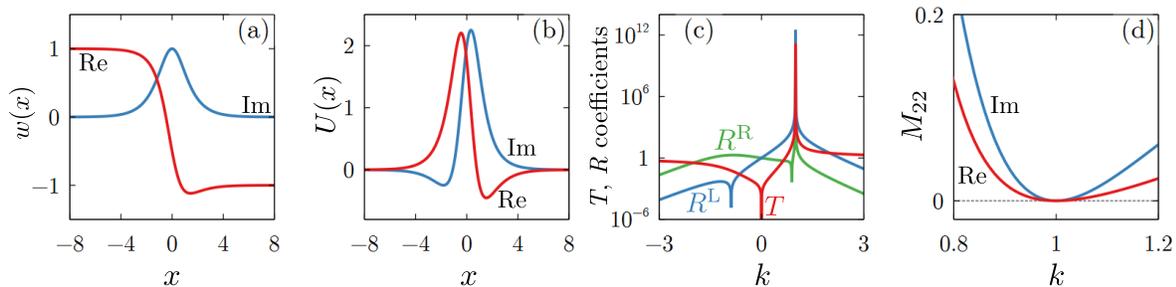}
	\caption{Potential (\ref{eq:tU}) with a second-order SS. (a,b) show function $\tilde{w}(x)$ and corresponding potential $\tilde{U}(x)$; real and blue curves correspond to real (Re) and imaginary (Im) parts, respectively. (c) and (d) show moduli of scattering coefficients and real and imaginary parts of transfer matrix element $M_{22}(k)$ (the latter are plotted only in the vicinity of the SS). In all panels $k_1=1$.}
	\label{fig:collision} 
\end{figure*}

As a simple example, consider (\ref{chi1}) with $a_0=k_1-k_2$ and $a_1=0$.
From (\ref{eq:tw}) we compute the limiting function $\tilde{w}(x)=-k_1\tanh x - \frac{1}{2}\sech x +ik_1\sech x$  which generates   the potential in the form
\begin{equation}
\label{eq:tU}
\tilde{U}(x) =   \left(2\left(k_1+\frac{i}{2}\right)^2(1+ i\sinh x) + \frac{1}{4}\right)\sech^2 x .
\end{equation}
This potential and   its scattering diagnostics are illustrated in figure~\ref{fig:collision}.  Both real and imaginary parts of the transfer matrix coefficient $M_{22}(k)$ remain nonnegative  in the neighborhood of the spectral singularity, which corroborates that the latter is of the second order. 

\section{Three  spectral singularities   and higher-order spectral singularities   in  pseudo-Hermitian Schr\"odinger operators} 
\label{sec:pseudo}

A particularly interesting situation emerges if one of the base functions $w_1$ or $w_2$   is real-valued. 
In this situation the potential $U(x)$, even if it is not $\PT$ symmetric, features a special antilinear symmetry \cite{Nixon} which enables  self-dual SSs \cite{ZezKon2020}. To be specific, if $w_1$ is real valued, then for all real $k$  the diagonal elements of the transfer matrix of the potential $U(x)$ satisfy the identities ~\cite{ZezKon2020}
\begin{equation}
\label{eq:M11M22}
M_{11}(k) = M_{22}^*(k)\frac{k+k_1}{k-k_1}, \quad M_{22}(k) = M_{11}^*(k)\frac{k-k_1}{k+k_1}.
\end{equation}
(If $w_2$ is real-valued, then in identities (\ref{eq:M11M22})  $k_1$ should be replaced with $k_2$).
Thus there is always a SS at  $k_1$. Moreover, using the identity    $M_{11}(k)=M_{22}(-k)$ which holds for all real $k$, we deduce that any eventual   SS at  $k$ different from $k_1$ is always self-dual.  Therefore, depending on the combination of $k_1$ and $k_2$, two different situations can be realized:
\begin{enumerate}
	\item   If $k_2=-k_1$, then there is a self-dual SS at $\pm k_1$. However, this SS is distinctively different from self-dual SSs in section~\ref{sec:sdss} because in the present case SS at $k_1$ is at least \emph{of the second order}, whereas the opposite spectral singularity at $-k_1$ is generically of the first order. Indeed, by construction $M_{22}(k_1)=0$ and $M_{22}(-k_1)=0$. 
	Hence $M_{11}(k_1)=0$, and, using relations (\ref{eq:M11M22}), we readily observe that $M_{22}(k)=O((k-k_1)^2)$ as ${k\to k_1}$.
	
	\item  If $k_2$ is different from $-k_1$, then   the potential $U(x)$ features  \emph{three} SSs, two of which correspond to the self-dual SS at $\pm k_2$, and the third one is at  $k_1$. (These SSs will be generically of the first order).
	
\end{enumerate}

In order to formulate   the condition  for   function $w_1$  to be   real-valued,  
%
%
we represent $\chi_1 = \rho(x)e^{i\phi(x)}$, where $\rho$ and $\phi$ are real functions,
and from Eq. (\ref{eq:ww1}) with $j=1$ we obtain that imaginary part of $w_1$ disappears if
\begin{equation}
\label{eq:sin}
\sin\phi = \frac{\rho'}{k_1^2-k_2^2 - \rho^2}.
\end{equation}
Thus, starting with some    function $\rho(x)$ which has asymptotic behavior $|\rho|\to |k_1-k_2|$ as $x\to\pm\infty$,  is  differentiable twice
and smooth enough to ensure $|\rho'/(k_2^2-k_1^2 + \rho^2)|\leq 1$ for all $x\in\mathbb{R}$,
one can, in principle, recover the corresponding phase $\phi$ and then obtain function $\chi_1(x)$ and real-valued function  $w_1(x)$. However, the procedure is not   completely straightforward,   because (\ref{eq:sin}) defines only sine of $\phi$ but does not ensure the required asymptotic behaviour of $\chi_1$ at the infinities. Meantime, the requirement for $\arg\chi_1$ to change from $0$ to $\pi$ may result
 in discontinuities of the first kind of $\chi_1$. In order to illustrate how these problems can be overcome, we explore the substitution for $\rho$ in the form of an odd function with a discontinuity of the second kind at $x=0$:  
%
\begin{equation}
\rho =  \frac{1}{x(x^2+1)} - (k_2-k_1) \tanh (ax).
\end{equation}
Here $x^2+1$ is incorporated in the denominator  in order  to speed up the algebraic decay at infinities, and $a>0$ is a free parameter. Considering the behavior around the discontinuity at the origin, using (\ref{eq:sin}) we observe that $\lim_{x\to 0}\sin\phi(x)=1$. Since $\rho(x)$ is an odd function of $x$, the required asymptotic behavior of $\chi_1(x)$ at the infinities is achieved if we fix   $\cos\phi=  \sqrt{1-\sin^2\phi}$.  As explained in section~\ref{sec:gen:constr}, for the resulting potential $U(x)$ to be well-behaved, in the vicinity of the discontinuity $\chi_{1}$ must behave as $i/x$, i.e., real part of $\chi_1$ must tend to zero as $x$ approaches the origin.  This leads to the conditions $\cos\phi(0)=d[ \cos\phi(0)]/dx=0$, i.e., $d^2[\sin\phi(0)]/dx^2=0$, which in its turn imposes an additional relation between  the wavenumbers $k_{1,2}$ and the free parameter $a$:
\begin{equation}
(k_1-k_2)(k_1+k_2-3a) + 3=0.
\end{equation}

\begin{figure}
	\centering
	\includegraphics[width=0.99\textwidth]{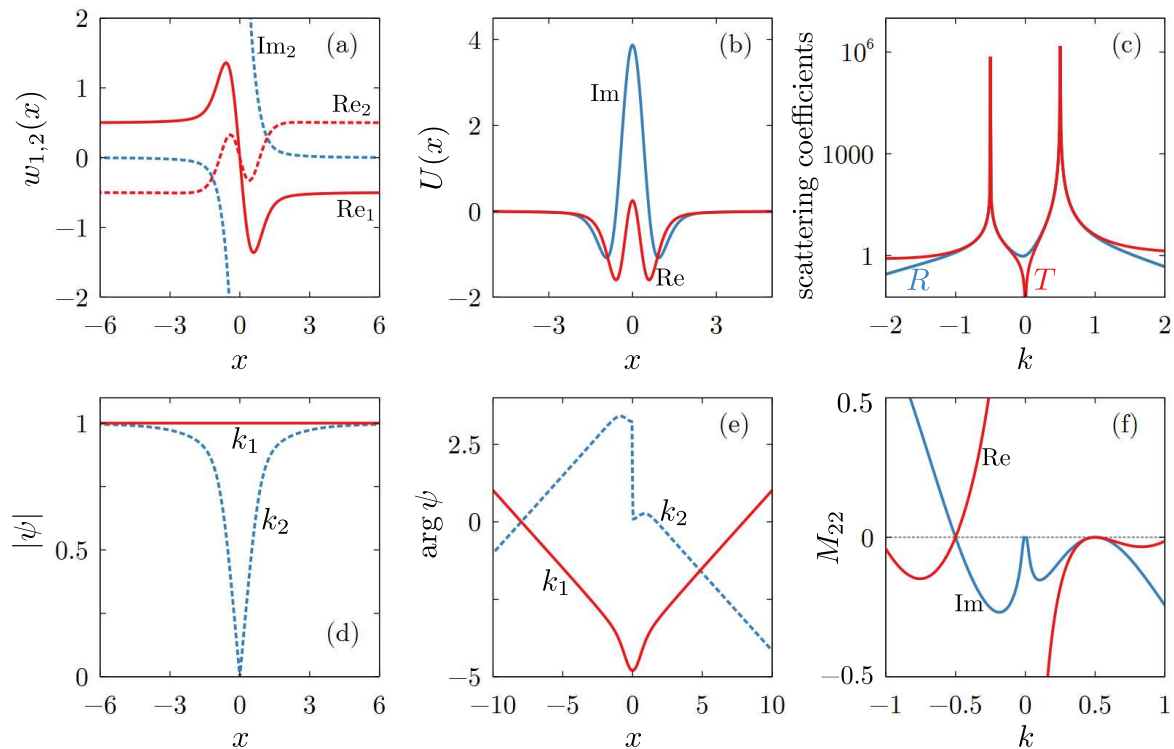}
	\caption{Example of a pseudo-Hermitian potential which features an ``unusual'' self-dual SS consisting of a first-order SS with $k>0$ and a second-order SS at $k<0$. (a) Real-valued function $w_1(x)$ and complex-valued function $w_2$ which generate  a complex-valued potential $U(x)$ are shown in (b).  Moduli  of transmission and reflection coefficients (c) and real and imaginary parts of the transfer matrix entry $M_{22}(k)$ (f). Since the potential is an even functions, moduli of the left and right reflection coefficients are equal. Panels (d,e) show amplitudes and unwrapped radian arguments (i.e.   phases) of the constant-amplitude laser solution corresponding to the SS   at   $k_1=1/2$ and the CPA solution  corresponding to the SS  at $k_2=-1/2$. }
	\label{fig:higher} 
\end{figure}

In order to illustrate  a higher-order SS [the above case (i)], we choose $a=1$ and $k_1=-k_2=1/2$. The obtained results are illustrated in Fig.~\ref{fig:higher}. In Fig.~\ref{fig:higher}(a) we observe that function $w_1$ is real-valued, whereas function $w_2$ has nontrivial real and  imaginary parts, the latter with  a discontinuity $i/x$. The potential $U(x)$, which  is an even, continuous and smooth function, is plotted in figure~\ref{fig:higher}(b). It  decays as $\propto x^{-3}$ as $x\to\pm\infty$. As shown in figure~\ref{fig:higher}(c), scattering coefficients feature sharp resonances at $\pm 1/2$ which is a manifestation of the self-dual SS. At the same time, plotting real and imaginary part of the transfer matrix coefficient $M_{22}(k)$ in figure~\ref{fig:higher}(f), we observe that spectral singularities at $k_1=1/2$ and $k_2=-1/2$ are distinctively different: the former is of the second order (real and imaginary parts of $M_{22}(k)$ remain negative in the vicinity of the SS, except for the very point of SS where they are both zero), and the latter is of the first order. In addition, in figure.~\ref{fig:higher}(d,e) we show amplitudes and phases of the lasing solution at $k_1$ and the CPA-solution at $k_2$. The laser solution is constant-amplitude due to the fact that $w_1$ is real valued, and the CPA solution has a zero at $x=0$ which agrees with the previous discussion on the singularity of function $w_2$.

\begin{figure*}
	\centering
	\includegraphics[width=0.99\textwidth]{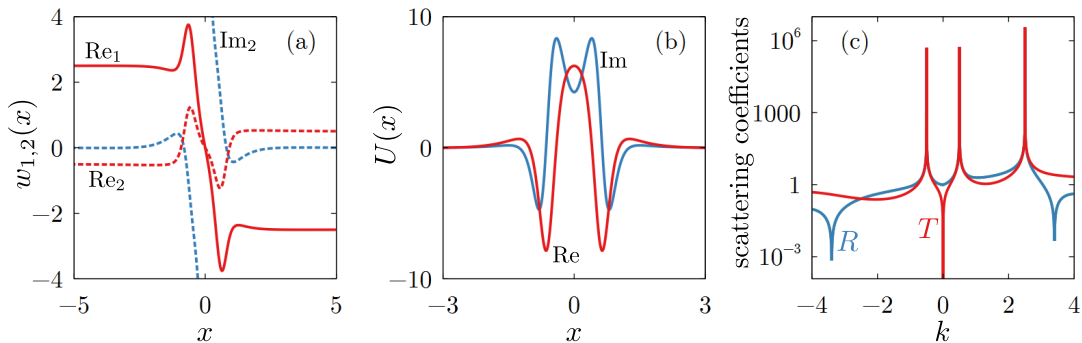}
	\caption{Example of a pseudo-Hermitian potential which features three SS, two of which form a self-dual SS. (a) Real-valued function $w_1(x)$ and complex-valued function $w_2$ which generate  a complex-valued potential $U(x)$ shown in (b).  (c) Moduli of the scattering coefficients;  the left and right reflection coefficients are equal. Three peaks at $k=\pm 1/2$ and $k=5/2$ correspond to a self-dual SS $k=\pm 1/2$ coexisting with a simple SS at $k=5/2$.}
	\label{fig:ss+sdss} 
\end{figure*}

In order to provide an explicit example for the case (ii) with coexisting self-dual and simple spectral singularities, we choose $a=1$,  $k_1=5/2$ and $k_2=-1/2$. The resulting functions $w_{1,2}$, localized  potential $U(x)$,  and moduli of the scattering coefficients are plotted in figure~\ref{fig:ss+sdss}.   According to the analytical prediction, the potential   features a spectral singularity (laser) at $k_1$ and a pair of spectral singularities at $\pm k_2$ which correspond  to  laser and CPA   modes. 


\section{Three independent spectral singularities}
\label{sec:three}

\begin{figure*}
	\centering
	\includegraphics[width=0.99\textwidth]{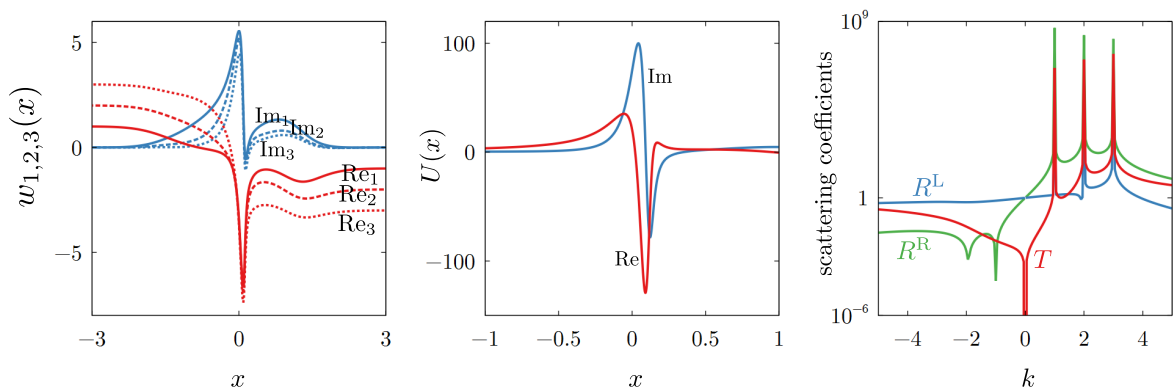}
	\caption{Example of a potential with three spectral singularities obtained according to our procedure using the substitution (\ref{eq:nu}) with $k_{1,2,3}=1,2,3$ and $z=-1/2-i/10$.  Panels (a) shows real (Re) and imaginary (Im) parts of base functions $w_{1,2,3}(x)$. Panel (b) shows real and imaginary parts of the resulting potential, and (c) shows the moduli of the scattering coefficients. }
	\label{fig:3ss} 
\end{figure*}

Now we address a general case where three spectral singularities are achieved at some  arbitrarily prescribed wavenumbers without any assumption of self-duality. According  to equation (\ref{eq:nu_chi}), the solution to this problem is reduced to solving a quadratic equation with respect to $\chi_1$ with the auxiliary function $\nu_1(x)$ chosen to satisfy the asymptotic behavior (\ref{asympt_nu}). Once $\chi_1(x)$ is known,  the base function $w_1$ and the full potential $U(x)$ can be determined.  The most challenging issue is that not any choice of $\nu_1$ leads to a ``suitable'' solution of the quadratic equation. For example, for a naive choice of $\nu_1$ in the form of a  purely real function or an anti-$\PT$-symmetric one, even chosen to satisfy asymptotic requirements   (\ref{asympt_nu}), the generated $\chi_1$ can be discontinuous or singular. 
Nevertheless, it is possible to make a suitable choice of $\nu_1$.  For an explicit example, let us choose $k_{1,2,3}=1,2,3$, i.e.  we look for a potential that lases at three different wavenumbers, and explore the following substitution which sufficiently rapidly approached the   asymptotic behavior (\ref{asympt_nu}): 
\begin{equation}
\label{eq:nu}
\nu_1 = \frac{k_{1}-k_{2}}{k_{2}-k_{3}}  + (ix+z)e^{-x^2},
\end{equation}
where  $z$ is a complex parameter. Three base functions  $w_{1,2,3}(x)$ obtained for $z=-1/2-i/10$ and the  resulting potential $U(x)$ are presented in figure~\ref{fig:3ss}(a,b). Plots of moduli  of the scattering coefficients in figure~\ref{fig:3ss}(c) clearly indicate  three coexisting resonant peaks which validate the result.


\section{Conclusion}
\label{sec:concl}

In this paper we have described an algorithmic approach for designing complex potential resulting in simultaneous emergence of two and three SSs in the spectra of the respective Schr\"{o}dinger operators. The construction is based on the universal form of the potential generated by a base function: different SS solutions are constructed using different base functions, while all these base functions  result in the same complex potential. Using this approach, we have been able to construct potentials that result in two and three spectral singularities. A procedure for construction of potentials that feature second-order spectral singularities has been also elaborated. In the situation, when the potential is pseudo-Hermitian, we have implemented a situation when one simple and one self-dual SS coexist in the spectrum, and when a self-dual SS consists of a first- and a second-order SSs. %
 In the meantime, a systematic procedure of obtaining complex potentials featuring any number of SSs in their spectra at different wavelengths remains an open question.

\ack

VVK is grateful to Z. Ahmed for the comments stimulated this work. VVK acknowledges financial support from the Portuguese Foundation for Science and Technology (FCT) under Contract no. UIDB/00618/2020. The work of DAZ   was supported by the Foundation for the Advancement of Theoretical Physics and Mathematics ``BASIS'' (Grant No.~19-1-3-41-1).

\end{document}